\documentclass[a4paper]{jpconf}
\usepackage{graphicx}
\usepackage{dcolumn}
\usepackage{epsfig,amsmath}
\graphicspath{{pics/}}    
\def\d{{\rm d}}
\def\be{\begin{equation}}
\def\ee{\end{equation}}

\begin{document}
\title{Dilepton production in proton-proton collisions at top RHIC energy}
\author{J Manninen$^1$, E L Bratkovskaya$^2$, W Cassing$^1$ and O Linnyk$^2$}
\address{$^1$ Institut f\"{u}r Theoretische Physik,
      Universit\"{a}t Giessen, 35392 Giessen, Germany}
\address{$^2$ Institut f\"{u}r Theoretische Physik,
      Universit\"{a}t Frankfurt, 60054 Frankfurt, Germany}

\ead{Mannisenjaakko@gmail.com}

\begin{abstract}
We study dielectron production in proton-proton collisions at top RHIC beam
energy within an extended statistical hadronization model. The invariant mass spectrum
of correlated dielectron pairs is evaluated in the low invariant mass region
and calculated results are compared with the PHENIX experiment. The model is 
found to be able to describe the data very well up to invariant masses of 
1 GeV with few adjustable parameters.
\end{abstract}


\section{Introduction}

Correlated dilepton radiation has been considered for a long time to 
be an important tool to study the hot and dense nuclear matter created
in relativistic nuclear collisions. The idea behind is that dileptons 
are emitted during all stages of the collision evolution and the 
electromagnetic cross sections with the created medium are believed to be 
small and so electromagnetic signals from the early stages might survive
the preceding evolution of the system. Thus, by studying the dilepton 
emission one can extract information also on the earlier, 
possibly exotic phases of the collision evolution. For this to hold true,
however, one has to carefully model the final (and dominant) stages of
the collision evolution where dilepton radiation is predominantly governed
by the decays of light mesons.   

The PHENIX collaboration has recently measured the 
correlated dielectron invariant mass spectrum in proton-proton [1] 
as well as in $Au+Au$ [2] collisions at the top RHIC beam energy. 
A large and so far un-explained excess of dielectrons in (semi)central 
heavy-ion collisions is seen in the data below the mass of $\phi$ meson 
when compared with various model calculations while the spectrum in 
$p+p$ collisions can be fairly well understood within the hadronic 
freeze-out cocktail calculations [1]. We present in this paper our 
baseline hadronic freeze-out cocktail calculation for the dielectron 
production in proton-proton collisions in the low invariant mass
region while the analysis of the heavy-ion collisions as well as
an extension to invariant masses larger than the $\phi$ meson mass
will be presented in a forthcoming publication [3].

\section{Statistical hadronization model}\label{SHM}

The statistical hadronization model (SHM) has been used successfully in
describing the hadronic yields and rapidity densities in high 
energy proton-proton collisions at SPS and RHIC [4-7].
In this work we employ the simplest version of the SHM, i.e. the 
analysis is performed in the grand-canonical ensemble, because calculations 
become considerably easier once we do not require
exact conservation of conserved quantities. The price to pay for the 
easier calculations is that one has to introduce a free parameter for 
each of the conserved quantities that are conserved in average only.

In the SHM, the primary hadron multiplicity of hadron type $i$
is calculated (omitting the fugacities) according
to
\begin{equation} \label{dndy}
N_i = V
\frac{2J_i+1}{(2\pi)^3}
\int \gamma_S^{n_s} \
\e^{-\sqrt{p^2 + m_i^2}/T} \ \d^3p.
\end{equation}
In Eq. (\ref{dndy}) $J_i$ denotes the spin, $p$ the three-momentum and $m_i$
the mass of the particle of the hadron species $i$.
In principle the model has 6 free parameters: $T$, $\mu_B$, $\mu_Q$, 
$\mu_S$, $\gamma_S$ and the system volume $V$, but we know that the central 
rapidity region is close to net charge 
free\footnote{estimates for the $\mu_B$ in the central rapidity region 
vary from around $20-30$ MeV in central $Au+Au$ collisions to around 10 
MeV in $p+p$ collisions ; $\mu_S\approx \mu_B/4$ and $\mu_Q\approx \mu_B/40$ 
are thus also small} and, furthermore, the dilepton radiation is 
dominated by decays of neutral mesons, whose multiplicities do not 
depend explicitly on the chemical potentials, and thus to a good approximation 
we can set the chemical potentials to zero. In this case we are left with 
3 free parameters ($T$, $\gamma_S$ and $V$) that describe the yields of all 
relevant hadrons. The $\gamma_S$ ($<1$) [8] parameter takes into account the 
empirical 
fact that the SHM tends to over-estimate the strange hadron yields in the 
elementary particle collision systems.

The SHM has been fitted to the STAR data measured in the central
rapidity region in $p+p$ collisions
[7] and we use the resulting fit parameters $T$=170 MeV and 
$\gamma_S$ = 0.6 in our analysis. We have re-adjusted the system volume 
such that the $\pi^0$ rapidity density at mid-rapidity is 
reproduced. All calculations are performed assuming that 
in each of the events there is only one large cluster produced. This picture 
is justified under special assumptions discussed for example in [9].
For the resonances with width larger than 2 MeV, Eq. (1) is convoluted
with the relativistic Breit-Wigner distribution in order to take into 
account the finite widths of the resonant states and the two dimensional 
integration over the momentum and mass is performed in order to calculate
the primary yields.

The evaluation of the dilepton emission with the SHM proceeds as follows:
First, we calculate the mean primary yields of different hadrons according 
to Eq. (1) and then - using these values - we sample event-by-event Poisson 
distributions for the primary yields of all hadrons and resonances.
Thus, even though the temperature, $\gamma_S$ and the  
system volume are kept the same for all events, we have somewhat different 
amounts of primary hadrons produced in each of the events and so the 
energy, for example, is not fixed but fluctuates from event to event. 

Next, unstable resonances decay into stable ones and as a final step, we 
let the vector mesons decay into dielectrons. The branching ratios of 
different hadrons and resonances into dileptons are typically very small and 
thus we take into account the dilepton radiation from the dominant sources 
only, which at RHIC energies are the low mass vector mesons. We do 
not consider the direct dilepton emission from most of the resonances in the 
cases where the 
contribution would be negligible. One should notice, however, that in our 
approach it is important to include all the known resonances ($\approx 300$)
because eventually they decay into the light mass vector 
mesons\footnote{for example about half of the 
$\rho^0$ mesons are produced primarily while the other half come from 
resonance decays}
and thus indirectly affect the dilepton yields.
The dilepton channels taken into account in our analysis are 
listed in Table 1.
\begin{table}[!ht]
\caption{Decay channels relevant for dielectron
production in $p+p$ collisions at $\sqrt{s}$=200 GeV. } 
\begin{center}
\begin{tabular}{llll}
\br
Hadron & direct & Dalitz & other \\
\mr
$\pi^0$ & & $\pi^0\rightarrow \gamma\, \e^+\e^-$ & \\
$\eta^0$ & & $\eta^0\rightarrow \gamma\, \e^+\e^-$ &
$\eta^0 \rightarrow \pi^+\pi^- \e^+\e^-$\\
$\eta'$ & & $\eta'\rightarrow \gamma\, \e^+\e^-$ &
$\eta' \rightarrow \pi^+\pi^- \e^+\e^-$\\
$\rho^0$ & $\rho^0\rightarrow \e^+\e^-$ & & \\
$\omega^0$ & $\omega^0\rightarrow \e^+\e^-$ & &
$\omega^0\rightarrow \pi^0\, \e^+\e^-$ \\
$\phi^0$ & $\phi^0\rightarrow \e^+\e^-$ & &
$\phi^0\rightarrow \eta\, \e^+\e^-$\\
\br
\end{tabular}
\end{center}
\end{table}

In general, the probability of a hadron to decay into a certain channel depends both on
the mass of the parent hadron as well as on the energies of the daughter particles. 
We have evaluated in detail the mass dependent partial widths into dileptons as well as 
the total widths of the vector mesons in this analysis while for the resonance decays, 
which do not directly produce dielectrons, we have taken into account trivial mass 
threshold effects only in our analysis. Formulae for the dielectron partial widths are 
known from the literature and we have used the standard expressions [10] while
the form factors used as well as a discussion on other details can be
found for example in [11].

The SHM is very efficient in counting the Lorentz invariant relative 
yields of different hadrons. This model, however, does not take into account the 
initial dynamical evolution of the colliding partons from the beam particles. 
Typically the SHM has been compared with experimental data which has been corrected 
for the limited acceptance in rapidity and transverse momentum, 
in which case the initial parton dynamics do not affect the final results. 
The PHENIX data has not been corrected for 
the acceptance effects and so in order to be able to compare our calculations with 
the data, we need to introduce an additional model that correctly accounts for the 
dynamics not taken into account in the SHM. We wish to do this as simply and 
transparently as possible so that we do not mask the main physics interest of this 
paper with complicated dynamical corrections. Thus, both in the longitudinal 
direction as well as in the transverse direction, we model the initial parton dynamics
by simply assuming a two dimensional Gaussian distribution for the momentum distribution 
of the clusters. Obviously, the means both in rapidity and $p_T$ are zero while we have 
fitted the widths of the clusters' $p_T$ and $y$ distribution such that the pion 
distributions become compatible with the measurements. 


\section{Dielectron production in the LMR in $p+p$ collisions at $\sqrt{s}$=200 GeV}

We compare our calculated dielectron invariant mass spectrum with the PHENIX
data [1] measured in proton-proton collisions at the top RHIC beam energy in
Figure 1. The contributions from the different vector meson decays are shown 
separately while the thick black line denotes the sum of all relevant contributions.
One can see that the model can do a fairly good job in describing the measured data 
in the LMR and it is clear that the aforementioned details like mass dependent 
partial widths must be taken carefully into account in order to describe the 
shape of the data
at all invariant masses. We have also added the background contribution from 
open charm decays (denoted with cc in the figure) but one can see that this 
contribution is negligible at all invariant masses except between the $\omega$
and $\phi$ meson peaks. We point out here that we have also taken into 
account the limited experimental mass resolution in the figure by running 
our calculated curves through a Gaussian smearing filter with a resolution of 10 MeV.
This way especially the $\phi$ meson peak and also the $\omega$ peak 
become broader and compatible with the experimental data. 

\begin{figure}[!ht]
\begin{center}
\includegraphics[angle=0,scale=1.0]{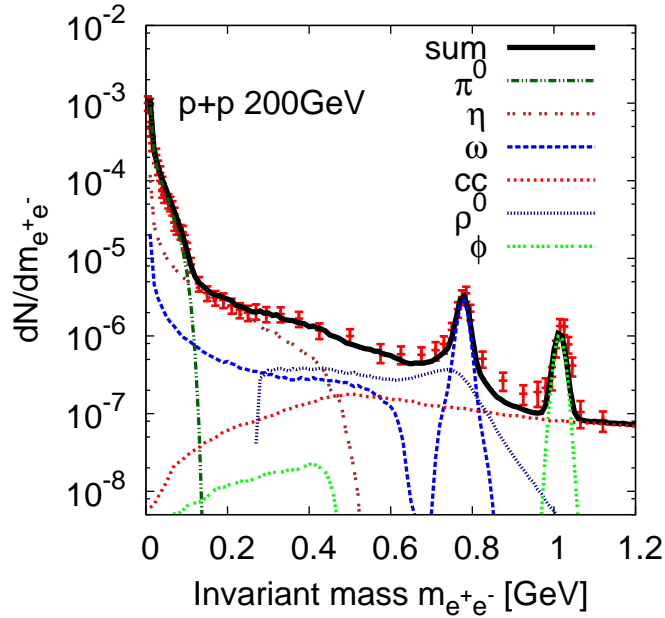}
\caption{Invariant mass spectrum of correlated pairs of electrons and
positrons in proton-proton collisions at $\sqrt{s}$ = 200 GeV. The
data are from the PHENIX collaboration~[1] while the
contribution from different dilepton emitting sources are
calculated as explained in the text. The full thick black line
denotes the sum from all relevant sources.}
\end{center}
\end{figure}

\section{Conclusions}

We have demonstrated that an extended statistical hadronization model can describe 
the dilepton radiation in ultra-relativistic proton-proton collisions  
at RHIC. In this work, we have implemented the statistical hadronization model
including most of the known resonances
and thus we have gone beyond the standard hadronic freeze-out 
cocktail calculations which typically include only a few different hadron and resonances 
species whose relative yields are taken either from the experiment or put in by hand.
In our approach, the relative yields of different hadron species and thus the relative
contribution at different invariant masses of the dielectron spectra is described 
with only 3 free parameters in the low invariant mass region. Thus, 
our work underlines the usefulness of the statistical hadronization model 
since together with simple and 
intuitive supplementary corrections for the initial state dynamics, 
we can address not only $4\pi$ integrated data but also the differential spectra 
measured by the PHENIX collaboration. 

\section*{Acknowledgments}
The authors would like to thank A. Toia for stimulating discussions.
Furthermore, E.L.B. and O.L. are grateful for financial support
from the 'HIC for FAIR' center of the 'LOEWE' program and J.M. for
support from DFG.

\end{document}